\documentclass[5p,times,twocolumn]{elsarticle}

\usepackage{amsmath,amssymb}
\biboptions{comma,sort&compress}
\usepackage[utf8]{inputenc}
\usepackage{graphicx}
\usepackage{bm}
\usepackage{hyperref}
\hypersetup{
colorlinks = true,
linkcolor = blue,
anchorcolor = blue,
citecolor = blue,
filecolor = blue,
urlcolor = blue
}
\usepackage[dvipsnames]{xcolor}

\graphicspath{{figs/}}

\newcommand{\music}{{\sc MUSIC}}
\newcommand{\isd}{{\sc iS3D}}
\newcommand{\urqmd}{{\sc UrQMD}}
\newcommand{\snn}{\sqrt{s_\mathrm{NN}}}
\newcommand{\etas}{\eta_s}
\newcommand{\mub}{\mu_B}
\newcommand{\dd}{{\rm d}}
\newcommand{\Tll}{T_{\ell\bar\ell}}
\newcommand{\llb}{\ell\bar\ell}
\newcommand{\Tgz}{T_{\gamma0}}
\newcommand{\Tg}{T_{\gamma}}
\newcommand{\Tavg}{\langle T\rangle}

\begin{document}
\begin{frontmatter}

\title{Multimessenger study of baryon-charged QCD matter in heavy-ion collisions}

\author[first,second,third]{Lipei Du}
\ead{ldu2@lbl.gov}
\affiliation[first]{organization={Department of Physics, McGill University},
            city={Montreal},
            postcode={H3A 2T8}, 
            state={Quebec},
            country={Canada}}
\affiliation[second]{organization={Department of Physics, University of California},
            city={Berkeley},
            postcode={94270}, 
            state={CA},
            country={USA}}
\affiliation[third]{organization={Nuclear Science Division, Lawrence Berkeley National Laboratory},
            city={Berkeley},
            postcode={94270}, 
            state={CA},
            country={USA}}
\date{\today}

\begin{abstract}
Multimessenger studies of heavy-ion collisions, using hadrons and electromagnetic probes, can reveal the properties of the created QCD matter from different perspectives. This study calculates the thermal dilepton invariant mass spectra and thermal photon transverse momentum spectra in Au+Au collisions at low beam energies from the Beam Energy Scan program, using a (3+1)-dimensional multistage hydrodynamic model calibrated by rapidity-dependent hadronic distributions. The effects of thermodynamic rapidity variation, baryon chemical potential, and fluid expansion on the spectra are explored. Methods for extracting temperature from photon and dilepton spectra are examined by comparison with the underlying hydrodynamic temperature. The possibility of combining photon and dilepton spectra to extract radial flow is also investigated. This study provides insights into measuring the thermodynamic properties of the created systems in heavy-ion collisions using multiple messengers through two fundamental interactions within the same framework.
\end{abstract}

\begin{keyword}
QCD matter \sep heavy-ion collisions \sep electromagnetic probes \sep multimessenger study
\end{keyword}

\end{frontmatter}

\section{Introduction}\label{sec:intro}

Understanding the many-body properties of quantum chromodynamics (QCD) and mapping its phase diagram remains a primary goal of the nuclear physics community \cite{Braun-Munzinger:2008szb,Sorensen:2023zkk}. Relativistic nuclear collisions \cite{Lee:1982qw,Shuryak:2014zxa} at various center-of-mass energies, such as those at the LHC and in the beam energy scan (BES) program at RHIC, are crucial for exploring the properties of QCD matter under extreme but controlled conditions, mapping a vast region of the QCD phase diagram \cite{Bzdak:2019pkr,An:2021wof,Du:2024wjm}. With the fixed target program concluded at RHIC, more experimental measurements will continue at facilities such as GSI/FAIR, HIAF, and NICA, probing the high baryon chemical potential regions of the phase diagram. These regions are also close to where neutron stars are located in the phase diagram, providing complementary physics information and exciting interdisciplinary opportunities \cite{Meszaros:2019xej,Dexheimer:2020zzs,Yao:2023yda}.

Systems created in heavy-ion collisions are highly dynamic and inhomogeneous, presenting significant challenges in extracting the thermodynamic properties of QCD matter from experimental measurements \cite{Bernhard:2019bmu,JETSCAPE:2020shq}. Despite challenges, the community has proposed various measures to directly extract these properties from experimental data \cite{Shuryak:2014zxa,Harris:2023tti}. Evaluating the effectiveness of these measures requires realistic simulations to test these measures and interpret the extracted properties, which are typically effective or average characteristics of the underlying QCD matter. For example, the freeze-out temperature and baryon chemical potential extracted by statistical thermal models \cite{Braun-Munzinger:2003pwq,Tawfik:2014eba,Andronic:2017pug} using identified hadron yields provide properties averaged over a rapidity window and average out inhomogeneity \cite{Du:2023gnv,Du:2023efk}.

Certain observables associated with electromagnetic probes, such as thermal photons and dileptons \cite{Peitzmann:2001mz,Rapp2013,Salabura:2020tou,Geurts:2022xmk}, can also provide thermodynamic measures. Although rare compared to hadrons in heavy-ion collisions, photons and dileptons interact through the electromagnetic interaction and thus remain unaltered once emitted. Thus, unlike hadrons, which carry information that can be easily distorted by dynamics, photons and dileptons carry information from their production point. Especially, those produced in the early stages of evolution can shed light on thermalization and chemical equilibration \cite{Gale:2021emg,Coquet:2021lca,Wu:2024pba}, processes that are still poorly understood \cite{Schlichting:2019abc,Berges:2020fwq}. However, electromagnetic probes are generated throughout the evolution of the collision fireball, where thermodynamic properties undergo significant spatial and temporal variations. Thus, the information extracted from measured electromagnetic observables is essentially integrated over these variations, representing effective or average characteristics of the underlying QCD system \cite{Peitzmann:2001mz,Rapp2013,Salabura:2020tou,Geurts:2022xmk}.

This study investigates the thermodynamic properties measured by hadrons, thermal dileptons, and thermal photons within the same framework, using a (3+1)-dimensional hydrodynamic description of baryon-charged QCD matter created at low center-of-mass energies of the BES program (Sec.~\ref{sec:model}). After examining the effects of thermodynamic rapidity variation, baryon chemical potential, and transverse expansion on the electromagnetic spectra (Sec.~\ref{sec:spectra}), this study explores the temperatures extracted from these different probes, aiming to understand their correlations during the evolution (Sec.~\ref{sec:temperature}). Additionally, the possibility of combining photon and dilepton spectra to extract radial flow is investigated (Sec.~\ref{sec:flow}). This multimessenger study of heavy-ion collisions provides insights into understanding these collisions through two fundamental interactions within the same framework, exploring how to extract the thermodynamic properties of the created systems using multiple experimental probes.

\section{Model and setup}\label{sec:model}

In this study, the dynamics of heavy-ion collisions are simulated using a (3+1)-dimensional multistage hydrodynamic model, incorporating \music~\cite{Schenke:2010nt,Schenke:2011bn,Paquet:2015lta}, \isd~\cite{McNelis:2019auj}, and \urqmd~\cite{Bass1998,Bleicher1999}. The simulation is initiated at a constant proper time $\tau_0$ with parametric 3-dimensional initial conditions. During the hydrodynamic stage, dissipative effects from the shear stress tensor and net baryon diffusion current \cite{Denicol2018,Du:2019obx,Du:2021zqz} are considered, while bulk viscous pressure is excluded. The particlization process, where hadrons are sampled from hydrodynamic fields, occurs on the freeze-out surface defined by a constant energy density $e_{\rm fo}=0.26\,$GeV/fm$^3$, the same value used in Refs.~\cite{Churchill:2023vpt,Churchill:2023zkk} for dilepton emission calculations. The same models have been implemented and described in Refs.~\cite{Churchill:2023vpt,Du:2023efk}, focusing on dilepton emission and hadron production at BES energies, respectively.

\begin{figure}[t]
    \centering 
    \includegraphics[width=0.33\textwidth]{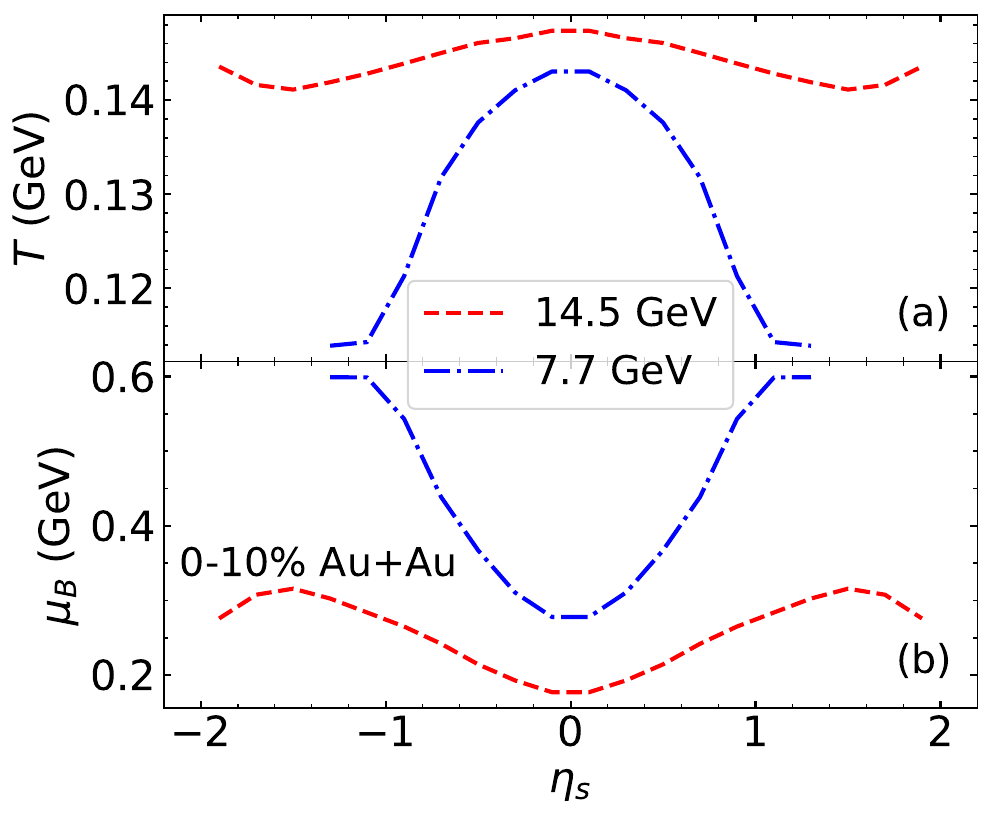}
    \caption{Energy density-weighted average of (a) temperature $T$ and (b) baryon chemical potential $\mub$ as a function of spacetime rapidity $\etas{\;=\;}\frac{1}{2}\ln\left(\frac{t+z}{t-z}\right)$ for fluid cells on the freeze-out surface, for 0--10\% Au+Au collisions at center-of-mass energies of 7.7 GeV (blue dot-dashed) and 14.5 GeV (red dashed).} 
    \label{fig:freezeout}%
\end{figure}

This study utilizes hadronic measurements at midrapidity by the STAR Collaboration \cite{STAR:2019vcp} and methods for constructing rapidity-dependent hadronic distributions not experimentally available \cite{Du:2023efk} to calibrate the bulk dynamics of Au+Au collisions at $\snn=14.5\,$GeV. This energy has not been considered in Refs.~\cite{Du:2022yok, Churchill:2023vpt,Churchill:2023zkk, Du:2023efk}, thus bridging the gap between collisions investigated at $\snn=7.7\,$GeV and 19.6 GeV. Details of the calibration at 14.5 GeV and the corresponding hadronic distributions are provided in \ref{app:bulk}. The agreement between the model calculations and experimental measurements at midrapidity, along with the constructed rapidity distributions for regions away from midrapidity, is verified for identified hadron yields (see Figs.~\ref{fig:rapidity_dist} and \ref{fig:BES_results}) to constrain thermodynamic properties and for the mean transverse momenta of identified hadrons to constrain the transverse flow of the bulk medium. 

The averaged thermodynamic properties, specifically temperature $\langle T\rangle$ and baryon chemical potential $\langle \mub\rangle$, at 14.5 GeV are shown in Fig.~\ref{fig:freezeout}. Using midrapidity values as a reference, Figure~\ref{fig:freezeout} shows that $\langle T\rangle$ can vary by approximately 4\% and $\langle \mub\rangle$ by approximately 80\% at 14.5 GeV, while at 7.7 GeV, $\langle T\rangle$ and $\langle \mub\rangle$ vary by approximately 25\% and 115\%, respectively. The thermodynamic properties exhibit weaker rapidity variation than at 7.7 GeV but moderately stronger than at 19.6 GeV \cite{Du:2023gnv,Du:2023efk}, showing a smooth transition between the 7.7 and 19.6 GeV collisions. 

Once the hydrodynamic model is calibrated by the hadronic observables, the entire spacetime evolution of all fluid cells is available. This study focuses on electromagnetic probes at the two low center-of-mass energies, $\snn{\;=\;}7.7$ and 14.5 GeV, where the rapidity variation in thermodynamic properties and effects of baryon chemical potential is expected to be the strongest among the energy range $\snn{\;=\;}$7.7--200 GeV. To calculate thermal photon ($\gamma$) and dilepton ($\llb$) spectra in the lab frame (LF), their emission rates \cite{Arnold2001ms,Ghiglieri2013,Ghiglieri2014,Jackson2019,Churchill:2023vpt} in the local rest frame (LRF), $\omega \dd^3\Gamma_\gamma/\dd^3\boldsymbol{k}$ and $\dd^4\Gamma_{\llb}/(\dd\omega \dd^3\boldsymbol{k})$, are convoluted with fluid cells over spacetime coordinates $x$:
\begin{equation}\label{eq:spectra}
    \omega'\frac{\dd^3N_\gamma}{\dd^3\boldsymbol{k'}}=\int \omega\frac{\dd^3\Gamma_\gamma}{\dd^3\boldsymbol{k}} \dd^4x\,,\quad \frac{\dd^4N_{\llb}}{\dd\omega'\dd^3\boldsymbol{k'}}=\int  \frac{\dd^4\Gamma_{\llb}}{\dd\omega \dd^3\boldsymbol{k}}\dd^4x\,.
\end{equation}
Here, the LF four-momentum ${K'}^\mu{\;=\;}(\omega', \boldsymbol{k'})$ is boosted to $K^\mu{\;=\;}(\omega, \boldsymbol{k})$ in the LRF of the fluid cell with four-velocity $u^\mu(x)$ before applying the LRF emission rates\footnote{The prime superscript of lab frame variables will be dropped for brevity.}; $\omega{\;=\;}|\boldsymbol{k}|$ for photons.
The emission rates used in this study account for $\mub$-dependence \cite{Gervais2012,Shen:2023aeg} but ignore viscous corrections \cite{Shen:2013cca,Vujanovic2013,Vujanovic2016}.

\begin{figure}[t]
    \centering 
    \includegraphics[width=0.85\linewidth]{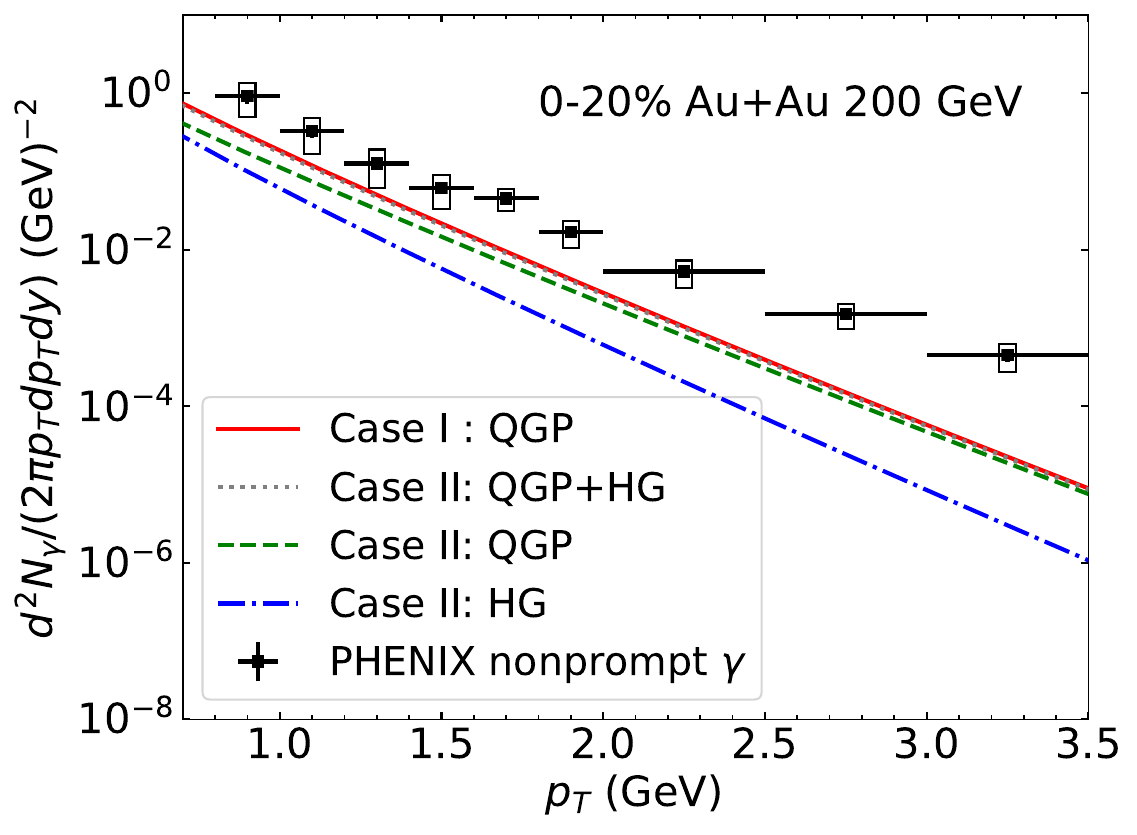}
    \caption{Thermal photon transverse momentum spectra for 0--20\% Au+Au collisions at 200 GeV in two scenarios. In Case I, the photon emission rate in the QGP is convoluted with hydrodynamic cells with temperatures above the Cleymans line \cite{Cleymans:2005xv}. In Case II, these cells are divided into two parts by temperature: higher temperatures are convoluted with QGP emission rates, and lower temperatures with hadronic gas (HG) rates. For Case II, the contributions from the QGP (green dashed) and HG (blue dot-dashed) are shown separately and combined (gray dotted); the gray dotted line is right on top of the red solid line. The nonprompt photon spectrum measured by the PHENIX Collaboration is also shown for comparison \cite{PHENIX:2008uif,PHENIX:2018for,PHENIX:2022rsx}.} 
    \label{fig:photon_200}%
\end{figure}

Fluid cells with temperatures above the Cleymans freeze-out line \cite{Cleymans:2005xv}, $T_C(\mub){\;=\;}a-b\mub^2-c\mub^4$, where $a{\;=\;}0.166\,$GeV, $b{\;=\;}0.139\,$GeV$^{-1}$, and $c{\;=\;}0.053\,$GeV$^{-3}$, are considered in the calculations. The line $T_C(\mu_B)$ closely aligns with the chemical freeze-out line extracted by the STAR Collaboration using identified hadron yields around midrapidity \cite{STAR:2017sal}. The spectra obtained are treated as thermal emissions from the Quark-Gluon Plasma (QGP) phase. The dilepton invariant mass spectra for Au+Au collisions at various center-of-mass energies have been systematically calculated and compared to STAR Collaboration measurements in Refs.~\cite{Churchill:2023vpt,Churchill:2023zkk}. Photon transverse momentum spectra for Au+Au collisions at $\snn{\;=\;}200\,$GeV are calculated and shown in Fig.~\ref{fig:photon_200} (Case I), where PHENIX Collaboration measurements are available for comparison \cite{PHENIX:2008uif,PHENIX:2018for,PHENIX:2022rsx}. The level of disagreement between the model and experimental data is consistent with previous studies \cite{Shen:2013vja,Paquet:2015lta}, and the observed discrepancies are well-known and remain under investigation. Nevertheless, it demonstrates that the current model calculation is in agreement with other similar theoretical calculations.

Additionally, another scenario dividing the contributions of considered fluid cells into QGP and hadronic gas (HG) contributions \cite{Shen:2013vja} is considered (Case II in Fig.~\ref{fig:photon_200}). Here, fluid cells above $T_C(\mu_B)$ are categorized into three groups by two more $T(\mu_B)$ lines, using the same parameterization as $T_C(\mu_B)$ but with $a=0.184$ and 0.22 GeV \cite{Shen:2013vja}, respectively. The thermal emission rate for QGP is applied to the group with higher temperatures, HG rates to lower temperatures, and the intermediate region is treated as a mixture of QGP and HG contributions. Figure~\ref{fig:photon_200} shows negligible differences between the final spectra under these two scenarios, and Case I is employed for the rest of the study.

\section{Results and discussion}\label{sec:results}
\subsection{Thermal dilepton and photon spectra}\label{sec:spectra}

\begin{figure}[tp]
    \centering 
    \includegraphics[width= 0.95\linewidth]{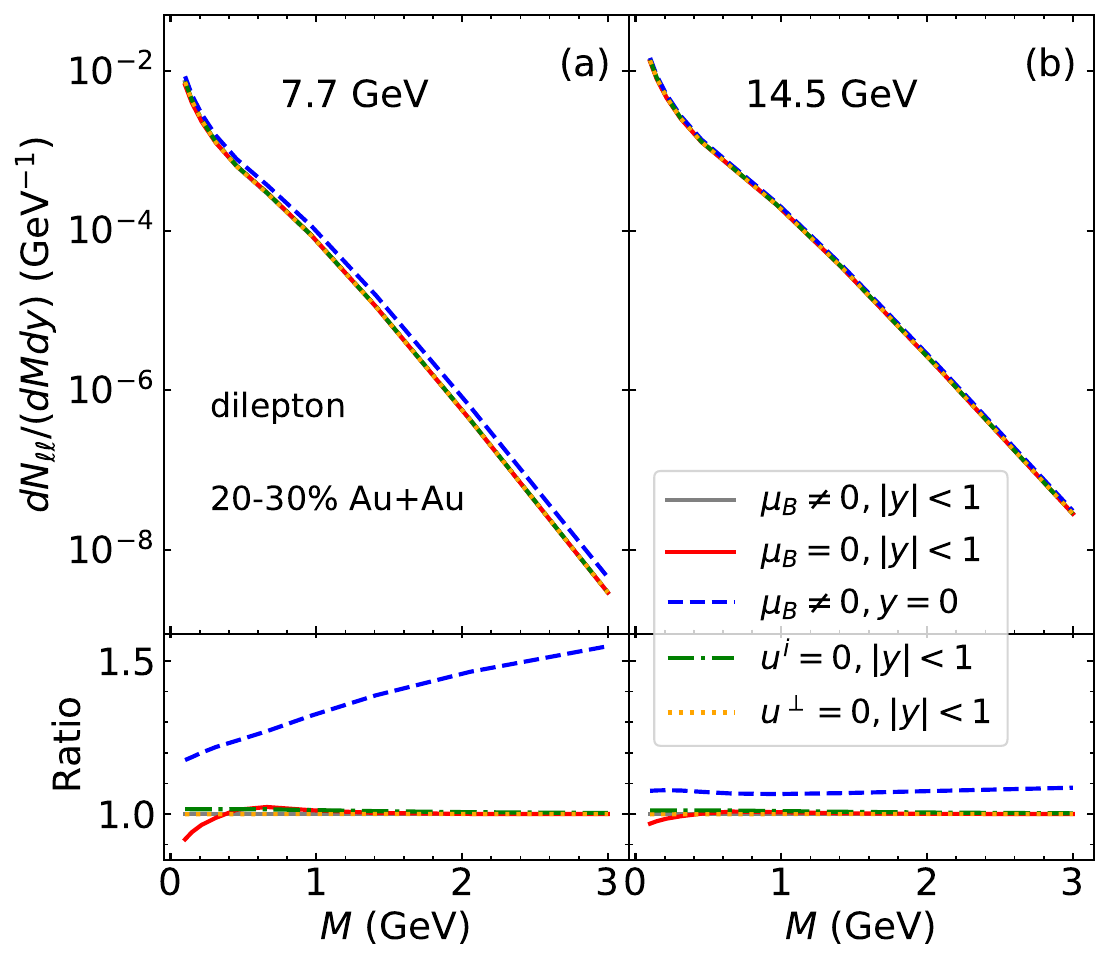}
    \caption{Thermal dilepton invariant mass spectra for 20--30\% Au+Au collisions at (a) 7.7 GeV and (b) 14.5 GeV, shown with various setups. The third case (blue dashed) is for midrapidity ($y=0$), while the others are for $|y|<1$. The second case (red solid) uses emission rates at $\mub=0$, while the others use nonzero $\mub$. The fourth (green dot-dashed) and fifth (orange dotted) cases illustrate scenarios with spatial flow components off ($u^i=0$) and transverse flow components off ($\boldsymbol{u}^\perp=0$), respectively. The lower panels show the ratios of these spectra, using the first case as the denominator.} 
    \label{fig:dilepton_setup}%
\end{figure}

\begin{figure}[tp]
    \centering 
    \includegraphics[width= 0.95\linewidth]{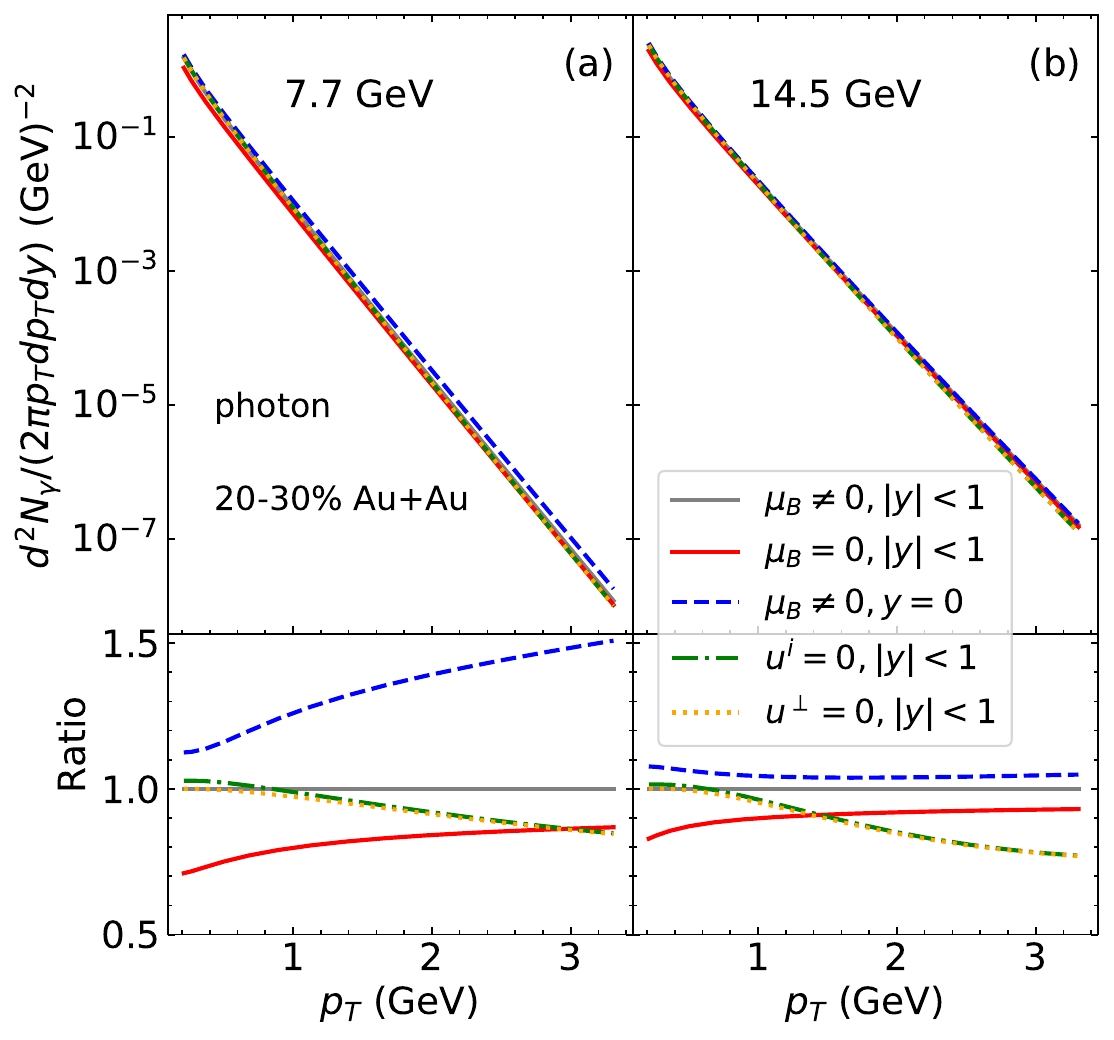}
    \caption{Thermal photon transverse momentum spectra for 20--30\% Au+Au collisions, shown with various setups, similar to Fig.~\ref{fig:dilepton_setup}.} 
    \label{fig:photon_setup}%
\end{figure}

This study focuses on the invariant mass ($M$) spectra of thermal dileptons and the transverse momentum ($p_T$) spectra of thermal photons, which are commonly measured in experiments \cite{STAR2015,STAR:2013pwb,STAR:2023wta,STAR:2015zal,HADES:2019auv,PHENIX:2008uif,PHENIX:2018for,PHENIX:2022rsx}. While theoretical calculations of transverse momentum spectra for dileptons are possible, measurements are typically challenging due to the low statistics of dilepton production in heavy-ion collisions. Various scenarios are considered in the calculation of these observables to investigate the effects of both thermodynamic and kinematic properties of the systems. The results for dileptons and photons are presented in Figs.~\ref{fig:dilepton_setup} and \ref{fig:photon_setup}, respectively, with a focus on intermediate central collisions in 20--30\% for clarity.

Figure \ref{fig:dilepton_setup} displays the dilepton invariant mass spectra for different setups. The first case, which includes $\mub$-dependence and considers dileptons within the rapidity window $|y|<1$, serves as the baseline. Comparing this with the second case, where $\mub$ is set to zero in the emission rates,\footnote{%
Note that $\mub$-dependence is only disabled in the emission rate, while the fluid cells contributing to the emissions remain unchanged.
} 
shows that the effect of $\mub$ is minimal, even at 7.7 GeV, which has the highest $\mub$ among the collision energies considered \cite{Churchill:2023vpt}. Comparing the third case, where the spectrum is calculated at midrapidity ($y=0$), with the baseline reveals a significant enhancement of the spectra: approximately 50\% at 7.7 GeV and about 10\% at 14.5 GeV at $M{\;=\;}3\,$GeV. As noted in Ref.~\cite{Churchill:2023vpt}, this substantial difference arises from the strong temperature variation in rapidity. The temperature is highest around midrapidity (as shown in Fig.~\ref{fig:freezeout}), leading to enhanced production compared to results within $|y|<1$, where the temperature is averaged to a lower value \cite{Du:2023efk}. This suggests that accurate calculations of dileptons at $\snn \lesssim 15\,$GeV require sophisticated (3+1)-dimensional modeling of the bulk medium evolution.

Dilepton invariant mass spectra are generally considered reference-frame independent, making them immune to distortions from the dynamical expansion of the systems, such as the Doppler effect. However, kinematic cuts in measurements, like $p_T$ range or rapidity window, can still influence the spectra if some dileptons are boosted out of or into the detector acceptance due to expansion. To investigate this, scenarios with spatial flow components off ($u^i=0$ with $i=x, y, \etas$) and transverse flow components off ($\boldsymbol{u}^\perp{\;=\;}0$) are also considered.\footnote{%
When some flow components are off, the temporal component $u^\tau$ is adjusted to ensure the four-velocity remains normalized.
}
The former additionally varies longitudinal expansion effects. Figure \ref{fig:dilepton_setup} shows that the Doppler effect on the dilepton invariant mass spectra is nonzero but negligible, at least when there is only a kinematic cut in rapidity, for the two center-of-mass energies considered.

Figure \ref{fig:photon_setup} presents photon transverse momentum spectra under the same four different setups explored for dileptons in Fig.~\ref{fig:dilepton_setup}. A much stronger $\mub$-dependence is evident \cite{Shen:2023aeg}, with zero baryon chemical potential significantly suppressing the spectra (by approximately 20\%) at lower center-of-mass energies, suggesting that photons can serve as a more effective probe of baryon evolution than dileptons. This $\mub$-dependence is more pronounced for softer photons at smaller $p_T$. The $\mub$-enhancement in photon production is primarily driven by the enhancement of quark bremsstrahlung \cite{Gervais:2012wd}. Additionally, the rapidity-dependence in the photon spectra is similar to that in the dilepton case. The variation seen in dilepton invariant mass spectra at small $M$ and photon transverse momentum spectra at small $p_T$ when comparing the scenarios at $y=0$ and $|y|<1$ in Figs.~\ref{fig:dilepton_setup} and \ref{fig:photon_setup} aligns with the variation in $\langle T\rangle$ along $\etas$ in Fig.~\ref{fig:freezeout} for both center-of-mass energies. 

The $p_T$ spectrum becomes softer when the flow is turned off \cite{Shen:2013vja,Paquet:2023bdx}, reducing by around 20\% at $p_T{\;\approx\;}3\,$GeV, with greater softening at 14.5 GeV due to turning off a stronger transverse flow. Interestingly, the results for the $u^i=0$ and $\boldsymbol{u}^\perp=0$ cases show a larger difference at small $p_T$ at 7.7 GeV, suggesting the effects of stronger longitudinal flow around midrapidity caused by a larger longitudinal pressure gradient, which has been observed in Ref.~\cite{Du:2023gnv}. Figure \ref{fig:photon_setup} indicates that the investigated thermodynamic and kinematic properties significantly impact the photon $p_T$ spectra, highlighting the necessity for sophisticated (3+1)-dimensional modeling of the bulk medium evolution for accurate photon spectrum calculations at such low collision energies. 

\subsection{Extraction of temperature}\label{sec:temperature}

After examining the effects on thermal dilepton and photon spectra, we now explore their effectiveness in measuring the temperature and radial flow of the QCD matter created in heavy-ion collisions. We focus on the spectra of dileptons and photons at midrapidity ($y=0$), dominated by emissions from fluid cells around $\eta_s{\;=\;}0$, for simplicity. Compared to the spectra within $|y|<1$ obtained in Sec.~\ref{sec:spectra}, the midrapidity spectra probe a more specific region of the system around $\etas=0$, and are thus less affected by thermodynamic rapidity variation and longitudinal flow.\footnote{%
It is useful to emphasize that the midrapidity spectra are integrated over the (3+1)-dimensional fluid extending along rapidity.
}
Correspondingly, we compare the properties of fluid cells within $|\eta_s|<0.1$ around mid-spacetime-rapidity.\footnote{%
It is important to note that the cells within $|\eta_s|<0.1$ (or at $\eta_s=0$) do not correspond exactly to those contributing to photon and dilepton production at midrapidity because of thermal smearing effects \cite{Churchill:2023vpt}; thus, no direct comparisons can be made.
}

\begin{figure}[t]
    \centering 
    \includegraphics[width= 0.95\linewidth]{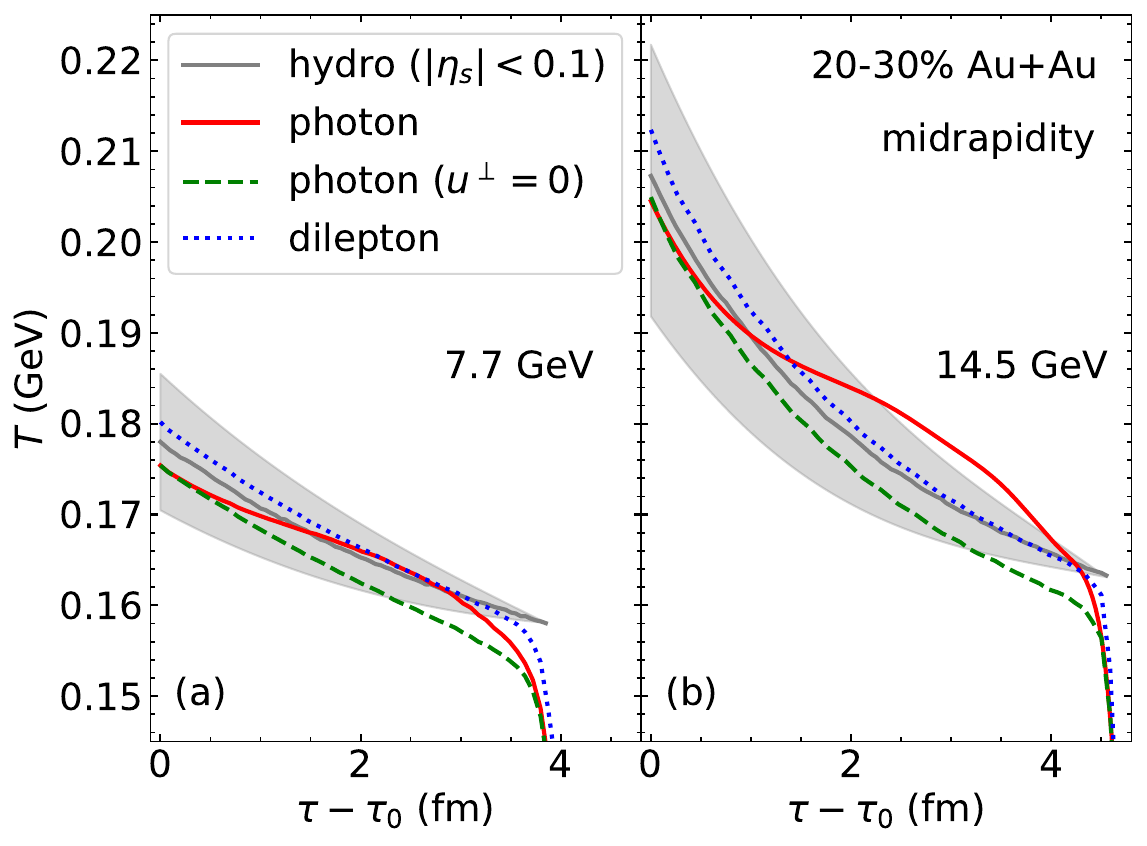}
    \caption{Comparison of temperature evolution at midrapidity for 20--30\% Au+Au collisions at (a) 7.7 GeV and (b) 14.5 GeV. The gray curves with bands represent the weighted average and variation of the temperature of hydrodynamic cells within $|\etas|<0.1$. Temperatures extracted from photon transverse momentum spectra are shown for cases with transverse flow (red solid) and without transverse flow (green dashed). The blue dotted line shows temperatures extracted from dilepton invariant mass spectra.} 
    \label{fig:temperature_extration}%
\end{figure}

To establish a baseline, we introduce a measure of the average thermodynamic properties of the underlying QCD fluid, considering its intrinsic inhomogeneity. Following Ref.~\cite{Churchill:2023zkk}, the average quantities are obtained by averaging over the properties of each fluid cell ($O$) weighted by its energy density ($e$) multiplied by the Lorentz boost factor ($\gamma{\;=\;}u^t$, with $u^t$ being the time component of the fluid four-velocity): $\langle O \rangle = \sum (O \cdot e \gamma)/\sum (e \gamma)$, where the summation is taken over all fluid cells within $|\eta_s|{\;<\;}0.1$ that are above $T_C(\mub)$. The weighting factor $e \gamma$ accounts for the fact that fluid cells with varying temperatures do not contribute equally to the observables; their contribution is proportional to their entropy or energy density. The variation of $O$ is obtained similarly by considering the deviation of $O$ at each fluid cell from the average $\langle O \rangle$.

The gray curves with bands in Fig.~\ref{fig:temperature_extration} illustrate the evolution of the weighted average ($\Tavg$) and variation of the temperature of hydrodynamic cells as a function of the shifted proper time $\tau-\tau_0$ \cite{Shen:2013vja,Churchill:2023zkk}. This temperature evolution, derived from hydrodynamic evolution calibrated by hadronic observables, represents information probed by hadrons. The figure reveals a much stronger temperature decrease during the hydrodynamic stage at 14.5 GeV, attributed to the system's stronger expansion. It also shows that the temperature variation diminishes and eventually shrinks to zero by the end of the evolution. This occurs because, as expansion continues and the fluid cell temperatures drop, fewer cells remain above $T_C(\mub)$; near the end of the evolution, only the region around the fireball center with $T_C(\mub)$ contributes to the average, causing the temperature variation to shrink to zero. Additionally, it is noteworthy that the temperature decrease slows down towards the end of the evolution. This slowdown happens because the expansion rate is lower near the fireball center, which will be illustrated by Fig.~\ref{fig:flow_extraction} below.

Both thermal dileptons and photons have been used as thermometers for QCD matter created in heavy-ion collisions \cite{Rapp:2014hha,Shen:2013vja}. To extract a temperature from their spectra, certain functional forms of their emission rates under approximations are employed. For thermal dileptons, the approximate large-$M$ (compared to $T$) behavior of the emission rate in Eq.~\eqref{eq:spectra} suggests that the integrated spectrum follows an approximate functional form ${\rm d}N_{\llb}/{\rm d}M {\;\propto\;} (M\Tll)^{3/2}\exp(-M/\Tll)$ in the mass region $1\, {\rm GeV}{\;<\;}M{\;<\;}3\, {\rm GeV}$ \cite{Rapp:2014hha,Churchill:2023vpt}. Similarly, for thermal photons, the spectra in the high-$p_T$  (compared to $T$) region are approximately exponential, ${\rm d}N_\gamma/({p_T\rm d}p_T) {\;\propto\;} \exp(-p_T/\Tg)$ \cite{Arnold2001ms,Shen:2013vja,Paquet:2023bdx}. By fitting their spectra in Figs.~\ref{fig:dilepton_setup} and \ref{fig:photon_setup} using these functional forms, one can obtain the effective or average temperatures $\Tll$ and $\Tg$. Compared to $\Tg$, $\Tll$ is typically considered immune to Doppler shift because the dilepton invariant mass spectra used for extraction are reference-frame independent, as confirmed in Sec.~\ref{sec:spectra}.

This study calculates the spectra at each time step and extracts time-differential temperatures for comparison with the hydrodynamic evolution. Figure \ref{fig:temperature_extration} shows that the dilepton temperature ($\Tll$), while slightly higher than the hydrodynamic average temperature, remains nearly parallel to the latter for most of the evolution, differing by about 1\%. $\Tll$ approaches and stays very close to $\Tavg$ in the final stage of the evolution. The photon temperature without transverse flow (thus no Doppler shift), denoted as $\Tgz$, exhibits similar behavior to $\Tll$, while slightly lower than $\Tavg$. Its offset from $\Tavg$ is also about 1\% for most of the evolution until the end. It's important to note that $\Tavg$ can be redefined with different weights, which could cause it to shift above $\Tll$ or below $\Tgz$. However, the key observation is that $\Tll$ consistently remains higher than $\Tgz$.\footnote{
Within a hydrodynamic spacetime evolution, the temperature extracted from dilepton spectra tends to be higher than the actual underlying temperature of the fluid cells, while the temperature extracted from photon spectra tends to be lower than the actual temperature, as shown in Refs.~\cite{Churchill:2023vpt} and \cite{Shen:2013vja}, respectively. The discrepancy between $\Tll$ and $\Tgz$ illustrated in Fig.~\ref{fig:temperature_extration} is also influenced by the different temperature dependencies of their emission rates. This occurs because, at each time step, the contributing fluid cells have varying temperatures, leading to different contributions to photon and dilepton spectra. A detailed exploration would be needed to quantitatively understand the offset between $\Tll$ and $\Tgz$.
}

Interestingly, $\Tll$ and $\Tgz$ evolve in parallel with a nearly constant offset, indicating that the temperature extraction from both photons (without Doppler shift) and dileptons is consistent. This suggests that $\Tll$ approaching $\Tavg$ at the final stage is coincidental, as $\Tgz$ does not approach $\Tavg$ in the same manner. In fact, the slowdown in the decrease of $\langle T\rangle$ towards the end of the evolution causes $\langle \Tll\rangle$ to approach $\langle T\rangle$. This is primarily due to the major contributing cells of $\langle T\rangle$ at the transverse center experiencing smaller transverse flow as already mentioned. The slowdown in the decrease of $\Tll$ and $\Tgz$ is weaker because the spectra at midrapidity used for temperature extraction receive contributions from cells away from midrapidity (i.e., those with $|\etas|{\;>\;}0.1$) where temperatures decrease faster due to both transverse and longitudinal expansions.

\subsection{Extraction of radial flow}\label{sec:flow}

Turning on the transverse flow, the photon temperature under the Doppler shift, $\Tg$, becomes a more practical measure accessible through experimental measurements \cite{PHENIX:2008uif,PHENIX:2018for,PHENIX:2022rsx}. Comparing $\Tg$ with $\Tgz$ reveals interesting effects of the transverse flow: $\Tg$ aligns with $\Tgz$ at the early stage of the evolution when transverse expansion is minimal. As the transverse expansion develops, $\Tg$ starts to deviate and becomes higher than $\Tgz$, reflecting a stronger Doppler shift \cite{Shen:2013vja}. This deviation is more pronounced at 14.5 GeV due to the hotter and more rapidly expanding fireball. Towards the end of the evolution, $\Tg$ approaches $\Tgz$ again because the contributing cells of the spectra above $T_C(\mub)$ are at the transverse center and experience less transverse flow.

\begin{figure}[t]
    \centering 
    \includegraphics[width= 0.95\linewidth]{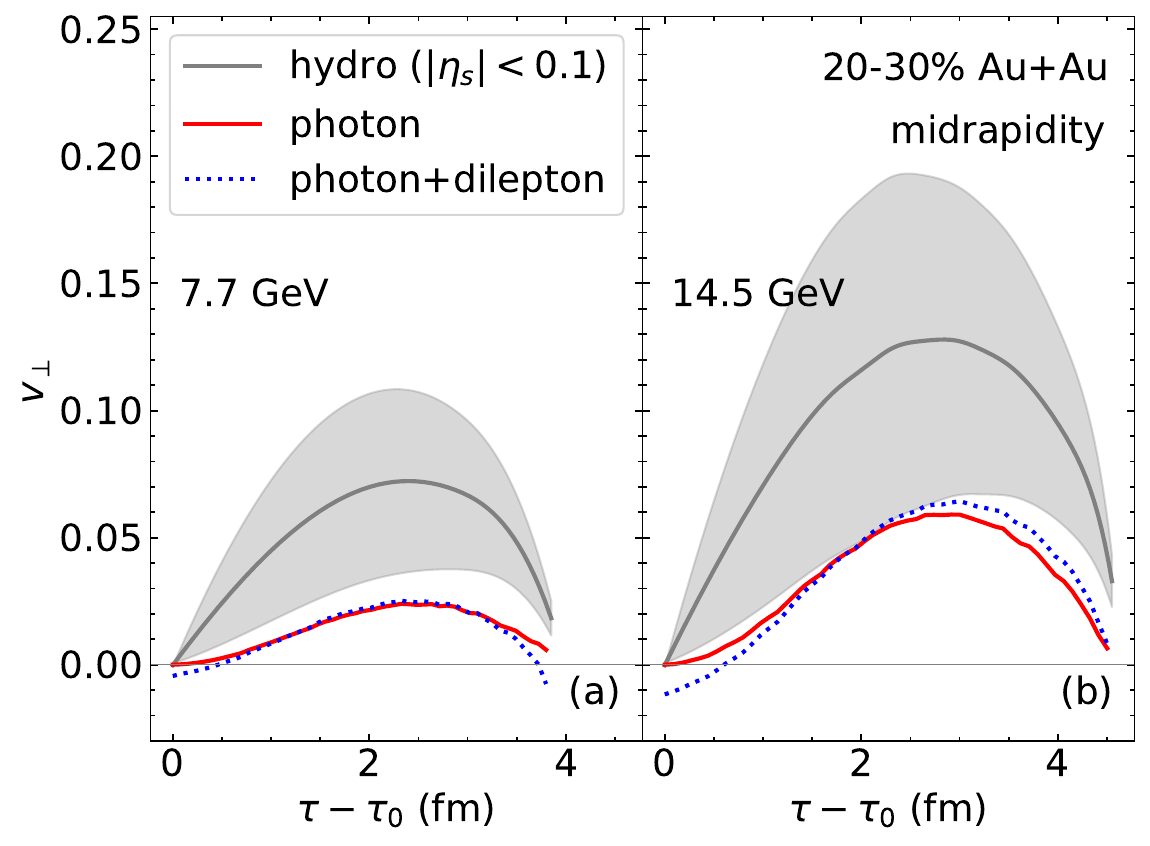}
    \caption{Comparison of extracted radial flow, similar to Fig.~\ref{fig:temperature_extration}. The red solid line shows radial flow combining temperatures from photon spectra with and without transverse flow in Fig.~\ref{fig:temperature_extration}. The blue dotted line shows radial flow combining temperatures from photon spectra with transverse flow and from dilepton spectra in Fig.~\ref{fig:temperature_extration}.} 
    \label{fig:flow_extraction}%
\end{figure}

Using temperatures extracted from photon spectra with and without transverse flow, it's possible to extract the radial flow $v_\perp$ with a blue shift factor \cite{feynman2015feynman}:
\begin{equation}\label{eq:flow}
    \Tg = \Tgz\sqrt{\frac{1+v_\perp}{1-v_\perp}}\,.
\end{equation}
Figure \ref{fig:flow_extraction} illustrates the extracted radial flow over time, compared with the hydrodynamic average radial flow $\langle v_\perp\rangle = \langle u^\perp/u^t\rangle$. The development and subsequent decrease of hydrodynamic radial flow near the end, due to the central fluid cells, are evident. Although the extracted radial flow is about half of the hydrodynamic value, it follows a similar trend throughout the evolution. The discrepancy arises mainly because central fluid cells, which dominate temperature extraction, experience smaller transverse flow, significantly weighting the extracted flow toward smaller values. Nonetheless, if the correlation between the extracted flow and the underlying hydrodynamic flow can be quantified through realistic simulations, the former can still be used to estimate the latter.

It's important to note that experimentally, photon spectra without Doppler shift are unavailable, and thus only $\Tg$ can be accessed from those with Doppler shift. While dilepton invariant mass spectra are immune to Doppler shift, the temperature $\Tll$ they extract does not match $\Tgz$, the temperature obtained from photon spectra without Doppler shift, as illustrated in Fig.~\ref{fig:temperature_extration}. However, since $\Tll$ and $\Tgz$ evolve in parallel with a nearly constant offset, by shifting $\Tll$ down by a constant (4 MeV at 7.7 GeV and 5 MeV at 14.5 GeV, based on Fig.~\ref{fig:temperature_extration}), we can estimate $\Tgz$. Combining this adjusted $\Tll$ with $\Tg$ to extract radial flow using Eq.~\eqref{eq:flow} yields results in good agreement with those obtained purely from photon spectra as shown in Fig.~\ref{fig:flow_extraction}. However, the downshifted $\Tll$ can sometimes be higher than $\Tg$, thus resulting in slightly negative flows.

Several points are noteworthy. First, time-differential spectra cannot be measured experimentally; instead, experiments measure spectra integrated over the entire evolution. Nonetheless, studying these spectra and their evolution using models provides valuable insights, such as identifying the correlation between $\Tll$ and $\Tgz$. Of course, extracting temperatures and radial flow should be further investigated using time-integrated spectra shown in Figs.~\ref{fig:dilepton_setup} and \ref{fig:photon_setup} to make experimental estimation feasible. Second, although the extracted temperatures and radial flow might deviate from underlying hydrodynamic quantities, they can still be useful in practice: if relationships are quantified between effective and underlying properties through realistic simulations, they can help interpret experimental measurements that provide effective properties \cite{Shen:2013vja,Churchill:2023zkk,Paquet:2023bdx}. Third, exploring different definitions of averaged hydrodynamic quantities $\langle O\rangle$ and identifying those matching extracted values will help understand connections between the production of different probes, including hadrons, photons, and dileptons. Investigating these points at various center-of-mass energies and centralities would provide a more comprehensive understanding, which is left for future work.

\section{Summary}\label{sec:summary}

Using a (3+1)-dimensional multistage hydrodynamic model calibrated with rapidity-dependent hadronic distributions, this study investigates electromagnetic probes, including dileptons and photons, in baryon-charged QCD matter created in heavy-ion collisions at two low center-of-mass energies (7.7 and 14.5 GeV) from the Beam Energy Scan program. The rapidity-dependent hadronic distributions are constructed for 14.5 GeV using universal scaling properties identified from available measurements across a wide range of center-of-mass energies. This calibrated model successfully describes experimental measurements for hadrons at midrapidity, indicating a smooth transition in thermodynamic properties as a function of beam energy from 7.7 to 200 GeV.

The study then explores the dilepton invariant mass spectra and photon transverse momentum spectra under various scenarios, focusing on the effects of thermodynamic rapidity variation, baryon chemical potential, and transverse flow of the underlying QCD matter. It is shown that thermodynamic rapidity variation significantly impacts both spectra, suggesting that sophisticated (3+1)-dimensional modeling is necessary for accurate calculations of electromagnetic production in collisions at such low energies. Additionally, while the dilepton spectra are insensitive to the baryon chemical potential and transverse flow, the photon spectra show sensitivity, indicating that photons could potentially serve as a better probe of baryon evolution.

In a multimessenger approach, the thermodynamic properties from the underlying hydrodynamic evolution (constrained by hadronic observables), dilepton spectra, and photon spectra are examined within the same framework, with the hydrodynamic properties serving as the baseline. Despite the complexities associated with inhomogeneity and dynamics of the QCD fireball, temperatures extracted from both dilepton spectra and photon spectra without Doppler shift are found to be consistent when intrinsic discrepancies are acknowledged. Although photon spectra without Doppler shift are not measurable, this study identifies a correlation between the temperatures extracted from photon and dilepton spectra and explores the possibility of combining measurable photon and dilepton spectra to extract radial flow.

In summary, this multimessenger study investigates hadronic and electromagnetic observables within the same framework to study baryon-charged QCD matter created in heavy-ion collisions at low center-of-mass energies. It investigates systems in a region of the QCD phase diagram close to neutron stars and uses multiple probes interacting through two fundamental interactions, similar to multimessenger astrophysics in neutron star research. By comparing the thermodynamic properties obtained from different types of observables, the study reveals the underlying QCD system from various perspectives. Identifying and examining correlations among these different types of observables in experiments can further help to validate the underlying theoretical aspects.

\section*{Acknowledgements}
The author acknowledges helpful conversations with Charles Gale and Greg Jackson. This work was partly supported by the Natural Sciences and Engineering Research Council of Canada. Computations were made on the computers managed by the Ohio Supercomputer Center \cite{OhioSupercomputerCenter1987} and on the B\'eluga supercomputer system from McGill University managed by Calcul Qu\'ebec and Digital Research Alliance of Canada.

\appendix

\section{Bulk properties at $\snn=14.5\,$GeV}\label{app:bulk}

This appendix details the calibration of bulk medium properties for Au+Au collisions at $\snn{\;=\;}14.5\,$GeV using the multistage model outlined in Sec.~\ref{sec:model}. The 14.5 GeV beam energy was previously missed in systematic studies of BES physics reported in Refs.~\cite{Churchill:2023vpt,Churchill:2023zkk,Du:2023efk}, due to experimental data being reported separately in Ref.~\cite{STAR:2019vcp} and overlooked. This study aims to bridge the gap between collisions at $\snn=7.7$ and 19.6 GeV by investigating the 14.5 GeV collisions.

\begin{figure}[bp]
    \centering 
    \includegraphics[width=0.3\textwidth]{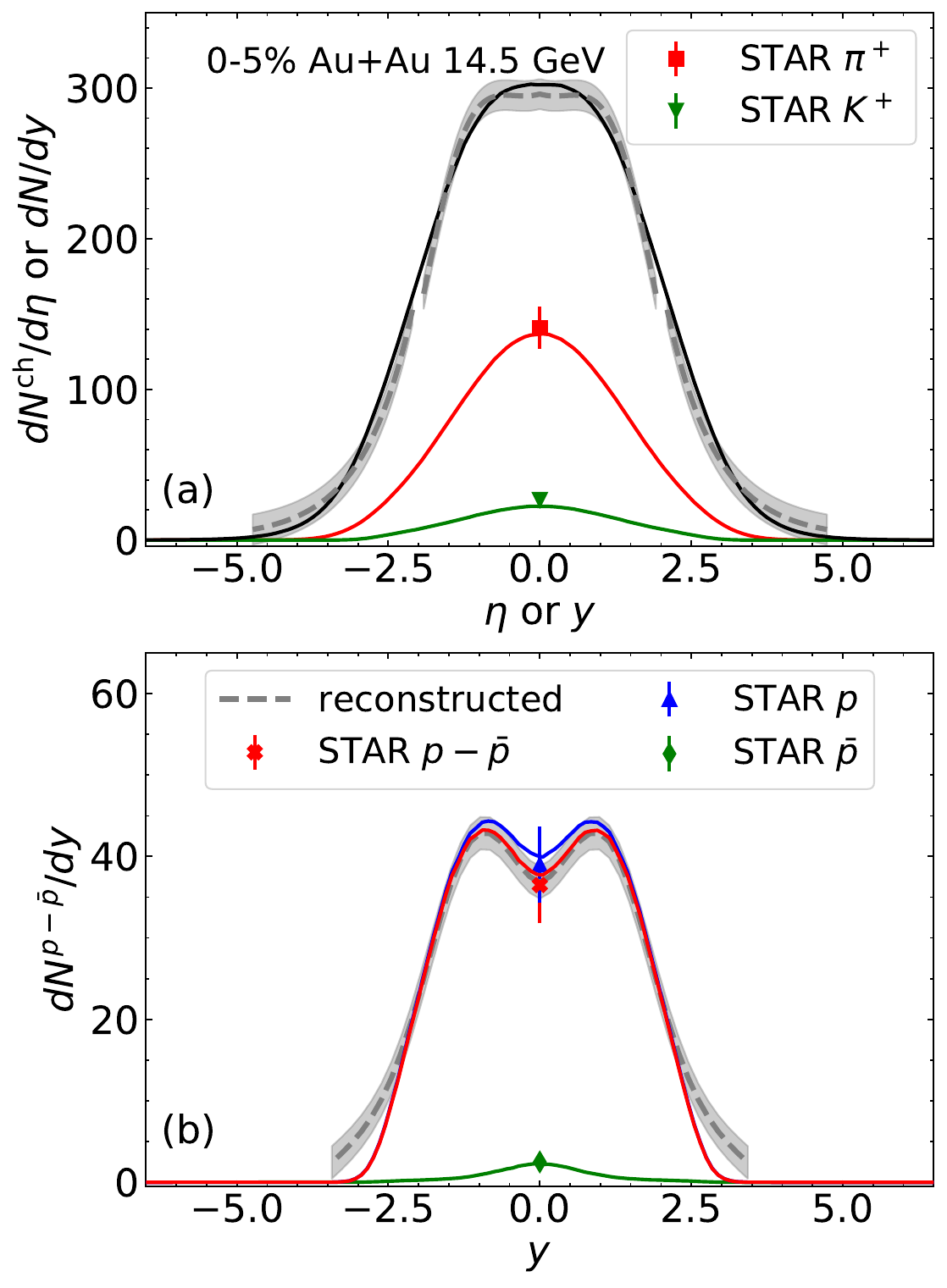}
    \caption{Rapidity-dependent distributions for 0--5\% Au+Au collisions at 14.5 GeV. The curves represent results from the calibrated multistage hydrodynamic model, while the markers indicate experimental measurements by the STAR Collaboration \cite{STAR:2019vcp}. The gray dashed lines with bands show constructed distributions for charged particle multiplicity $dN^{\rm ch}/d\eta$ in (a) and net-proton yields $dN^{p-\bar p}/dy$ in (b) using the methods from Ref.~\cite{Du:2023efk}.} 
    \label{fig:rapidity_dist}%
\end{figure}

\begin{figure}[htpb]
    \centering 
    \includegraphics[width=0.3\textwidth]{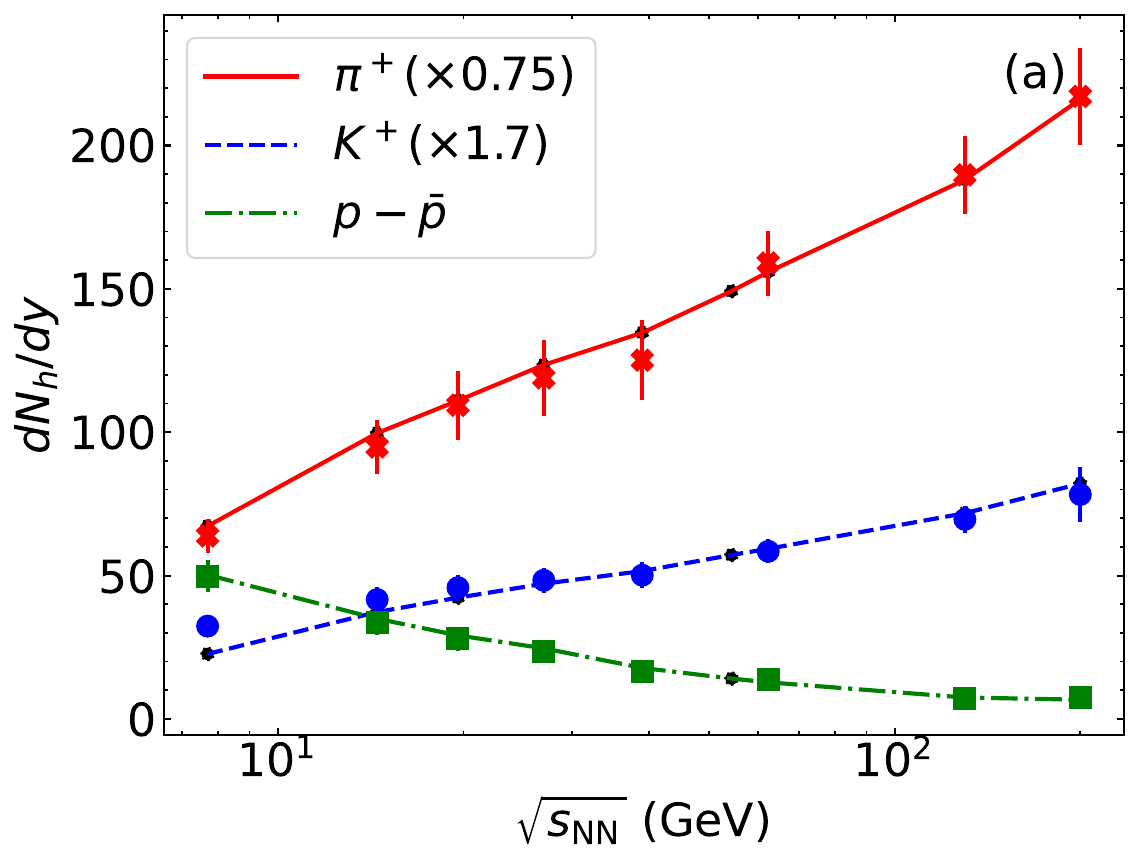}
    \includegraphics[width=0.3\textwidth]{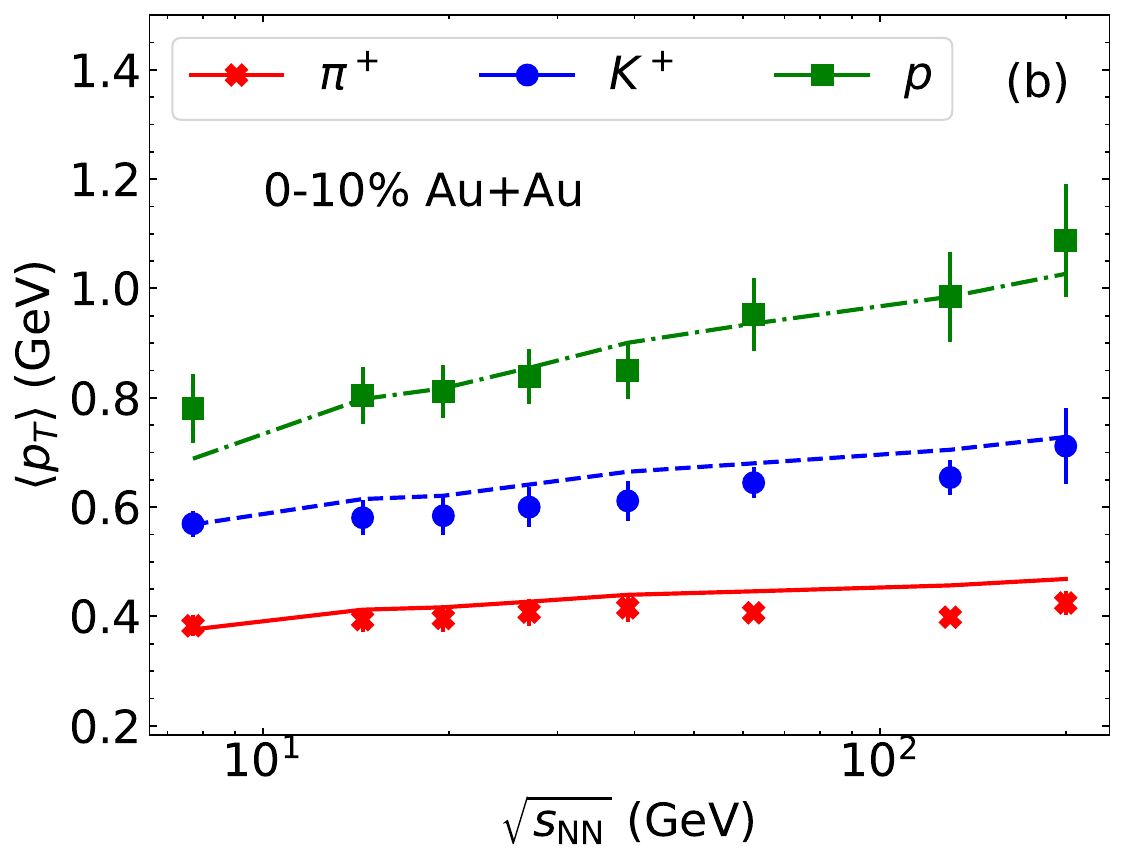}
    \caption{Identified hadron yields (a) and mean transverse momentum (b) around midrapidity for 0--10\% Au+Au collisions at nine center-of-mass energies ($\snn$) from 7.7 GeV to 200 GeV. Note that except for the 14.5 GeV results, those at other center-of-mass energies are taken from Refs.~\cite{Churchill:2023vpt,Du:2023efk}. The curves are results from the calibrated multistage hydrodynamic model, and the markers represent experimental measurements by the STAR Collaboration \cite{STAR:2008med,STAR:2017sal,STAR:2019vcp}.} 
    \label{fig:BES_results}%
\end{figure}

Same as hadron yield measurements at other center-of-mass energies by the STAR Collaboration, the data at $\snn=14.5\,$GeV were measured at midrapidity, posing challenges for calibrating bulk medium dynamics along the beam direction. This study employs the approach proposed in Ref.~\cite{Du:2023efk}, constructing rapidity-dependent hadronic distributions using universal scaling properties identified from a wide range of center-of-mass energies. This construction is applied to charged particle multiplicity $dN^{\rm ch}/d\eta$ and net-proton yields $dN^{p-\bar p}/dy$, shown in Fig.~\ref{fig:rapidity_dist}, to constrain bulk medium dynamics.

The model then generates identified hadron yields and mean transverse momenta, which are compared to available measurements around midrapidity, as shown in Figs.~\ref{fig:rapidity_dist} and \ref{fig:BES_results}. The good agreement observed validates the model and the method of constructing rapidity-dependent hadronic distributions as proposed in Ref.~\cite{Du:2023efk}. The results across center-of-mass energies from 7.7 GeV to 200 GeV, shown in Fig.~\ref{fig:BES_results}, suggest no dramatic changes in bulk medium properties, indicating a smooth transition in thermodynamic properties with beam energy from 7.7 to 200 GeV.

The average thermodynamic properties, such as temperature $\langle T\rangle$ and baryon chemical potential $\langle \mub\rangle$, as functions of spacetime rapidity $\etas$, are shown in Fig.~\ref{fig:freezeout} for 7.7 and 14.5 GeV, the two center-of-mass energies studied for electromagnetic probes in the main text. Fig.~\ref{fig:freezeout} indicates that $\langle T\rangle$ and $\langle \mub\rangle$ exhibit W-shaped and M-shaped distributions in $\etas$ at 14.5 GeV, features observed in Ref.~\cite{Du:2023gnv} only for results at 19.6 GeV among the four investigated center-of-mass energies (7.7, 19.6, 62.4, and 200 GeV). It is useful to note that the results for $T$ and $\mub$ in Ref.~\cite{Du:2023gnv} were obtained for values at the transverse center of the fireball, whereas here they represent weighted averaged properties over the transverse plane. Despite this difference, both methods yield similar results since the average properties are dominated by fluid cells at the transverse center, where temperatures are highest.

\bibliographystyle{elsarticle-num} 
\bibliography{refs}

\end{document}